\documentclass[aip,jcp,reprint]{revtex4-1}
\usepackage{graphicx}
\usepackage{amsmath}
\usepackage{bm}
\usepackage{amssymb} 
\usepackage[usenames, dvipsnames]{color}
\DeclareMathOperator{\erfc}{erfc}
 
\begin{document}
	
	\title{Donnan Grand Canonical Monte Carlo Simulations with Fast Ewald Summation for Calculating Individual Ionic Chemical Potentials}
	
	\author{Alexandre P. dos Santos}
	\affiliation{Instituto de F\'isica, Universidade Federal do Rio Grande do Sul, Caixa Postal 15051, CEP 91501-970, Porto Alegre, RS, Brazil.}
	
	\author{Amin Bakhshandeh}
	\affiliation{Escola de Qu\'imica e Alimentos, Universidade Federal do Rio Grande, Rio Grande 96203-900, Brazil}
	
	\author{Yan Levin}
	\email{levin@if.ufrgs.br}
	\affiliation{Instituto de F\'isica, Universidade Federal do Rio Grande do Sul, Caixa Postal 15051, CEP 91501-970, Porto Alegre, RS, Brazil.}
	
	\begin{abstract}
		We present a Donnan Grand Canonical Monte Carlo (GCMC) simulation method using fast Ewald summation to accurately and efficiently calculate the individual ionic excess chemical potentials of $\alpha_+:\alpha_-$ electrolyte solutions. By introducing a dynamic inner electrostatic potential that acts to penalize net charge fluctuations and maintain average charge neutrality ($\langle Q_t \rangle = 0$), the method enables single-ion GCMC trial moves without sacrificing convergence speed or breaking thermodynamic consistency. Unlike the Widom particle insertion method---which suffers from severe phase-space sampling bottlenecks and requires separate virtual insertions for each ionic species---the Donnan GCMC protocol simultaneously yields the individual excess chemical potentials $\mu_+^{\text{ex}}$ and $\mu_-^{\text{ex}}$ corresponding to a resulting bulk concentration $c_b$ with rapid numerical convergence. Varying the reference activity parameter $c_r$ allows us to efficiently construct the complete $\mu_\pm^{\text{ex}}(c_b)$ curves. Comparison with the Mean Spherical Approximation (MSA) demonstrates excellent agreement for size-symmetric electrolytes, while highlighting the failure of linear-response theories for size-asymmetric ions where electrostatic correlations and cavity effects become strongly coupled.
	\end{abstract}
	
	\maketitle
	
	\section{Introduction}
	
	The thermodynamic characterization of individual ionic species in electrolyte solutions remains a cornerstone of physical chemistry and electrochemistry~\cite{browne2018electrochemistry,levin2002electrostatic,levin2023new,newman2021electrochemical,devgan2025principles}. While only globally charge-neutral combinations of chemical potentials~\cite{consistent} are directly accessible through macroscopic thermodynamic measurements, single-ion properties govern a vast array of fundamental physical phenomena. These include ion-specific adsorption at fluid and solid interfaces, biological membrane transport, protein solubility and crystallization (the Hofmeister series), colloidal stability, and differential ion selectivity in targeted separation processes~\cite{PhysRevLett.103.257802,Beyer,beyer2025correct}.
	
	The necessity of working with individual ionic chemical potentials becomes particularly acute when investigating confined aqueous electrolytes, such as those found in nanoporous supercapacitor electrodes, biological channels, clay minerals, and nano-fluidic devices~\cite{PhysRevResearch.5.043174,Boda2,Guggenheim,Stenberg2020,Aslyamov2022,Boda2011,Hsiao2022,Hedstrom2012,Aluru2023}. In macroscopically open or confined geometries in contact with an external electrolyte reservoir, strict local charge neutrality can break down, leading to significant charge accumulation or depletion within the pore matrix~\cite{colla2016charge}. Because local species partitioning depends on the individual electrochemical potentials $\tilde{\mu}_+ = \mu_+ + q_+ \psi$ and $\tilde{\mu}_- = \mu_- + q_- \psi$, traditional modeling techniques that enforce local pair-wise insertion fall short~\cite{Malasics2010,Sorensen1992,deLeeuw1980}. To study such non-neutral confined environments using standard neutral-cluster insertion moves, one is forced to model the complete electrochemical cell---including the porous anode, the porous cathode, and a sufficiently large intervening bulk electrolyte reservoir---demanding massive simulation domains and leading to prohibitive computational costs~\cite{Boda3,PhysRevE.86.016703,Zeng2021}. Furthermore, multi-particle trial insertions suffer from exponentially vanishing acceptance rates at moderate-to-high ionic strength or inside sterically congested nanopores~\cite{Hatch,Kim}.
	
	A prominent example of this distinction arises in constant-pH and reactive electrochemical systems~\cite{landsgesell2019simulations}. The activity of hydronium ions---and thus the physical pH---is defined strictly in terms of the purely chemical (intrinsic) contribution $\mu_{\text{H}^+}$ of the electrochemical potential~\cite{labbez2006new,israelachvili2011,Beyer,Guggenheim,Atkins2023}. When a nano-confined geometry or porous medium develops an electrostatic inner potential difference relative to an external bulk reservoir, the local chemical potential shifts. Consequently, the local equilibrium pH inside the pore differs fundamentally from that of the bulk reservoir by a factor proportional to the Donnan potential~\cite{landsgesell2019simulations,Levincomment}. Accurately capturing such local reactive equilibria requires an explicit, decoupled knowledge of the single-ion chemical potentials.
	
	From a computational perspective, Grand Canonical Monte Carlo (GCMC) simulations offer a natural framework for studying open systems by allowing particle numbers to fluctuate at fixed chemical potential~\cite{Frenkel2002,Valleau1980}. However, executing GCMC for ionic fluids presents severe technical challenges. To avoid divergent electrostatic energy terms associated with net uncompensated charge in periodic boundary conditions, traditional GCMC simulations resort to the simultaneous insertion or deletion of neutral ion clusters ($\nu_+ \text{cations} + \nu_- \text{anions}$).
	
	To circumvent the computational overhead of simulating complete electrochemical cells, iterative GCMC algorithms have been introduced to dynamically adjust input chemical potentials on the fly to match target bulk concentrations~\cite{boda}. However, these original formulations relied on the minimum image convention paired with a simple truncated real-space Coulomb potential. While computationally straightforward, truncating $1/r$ electrostatic interactions is fundamentally problematic for ionic fluids. It introduces severe step discontinuities at the cutoff boundary, violates local charge-neutrality sum rules (such as the Stillinger–Lovett conditions), neglects the long-range screening tails essential for collective polarization, and completely misses the macroscopic boundary potential drops inherent to periodic systems~\cite{Fennell,Stillinger,Ballenegger,Kale2011,Falc}.
	
	To bypass these limitations and perform true single-ion GCMC trial insertions directly within a single region of interest, one requires an exact framework that operates under long-range periodic boundary conditions. Individual ionic chemical potentials can be computed using single-particle Widom test-particle insertion schemes, which properly account for the residual background self-energy in non-neutral Ewald systems~\cite{bakhshandeh2022widom} and remain the gold standard against which alternative calculation methods are benchmarked. Nevertheless, because Widom insertion relies on virtual test-particle insertions into unperturbed equilibrium configurations, calculating free energies in dense, strongly correlated, or size-asymmetric electrolytes requires extensive phase-space sampling, resulting in slow numerical convergence~\cite{Widom1963}.
	
	To overcome these sampling bottlenecks while maintaining exact electrostatics, a dynamic Donnan potential framework was recently introduced in the context of colloidal suspensions~\cite{bakhshandeh2022reactive} and reactive constant-pH Ewald simulations~\cite{levin2023new} of confined systems~\cite{levin2026modified}. By dynamically adjusting an electrostatic potential parameter to account for the Donnan potential drop between the system and the reservoir, this framework established the exact equivalence of electrochemical potentials between the system and the reservoir, directly predicting local pH shifts and charge regulation equilibria without requiring one to explicitly simulate the reservoir.
	
	In this work, we leverage this dynamic Donnan concept to formulate a Grand Canonical Monte Carlo methodology to directly calculate the individual ionic excess chemical potentials. By introducing an adaptive inner electrostatic potential $\psi_R$ that dynamically penalizes net charge fluctuations and enforces macroscopic charge neutrality ($\langle Q_t \rangle = 0$), our protocol executes single-ion GCMC insertion and deletion trial moves directly within a single simulation box. The framework uses the exact fast long-range 3D Ewald summation, which preserves the correct electrostatics, while explicitly accounting for the instantaneous Bethe potential $\phi_B$ associated with the boundary termination of the periodic Ewald crystal. For a given reference activity $c_r$, a single simulation run rapidly converges to yield both the physical bulk concentration $c_b$ and the individual excess chemical potentials $\mu_+^{\text{ex}}$ and $\mu_-^{\text{ex}}$ simultaneously. We validate the method against direct single-particle Widom insertion calculations as well as the Mean Spherical Approximation (MSA)~\cite{ho1988mean} for size-symmetric electrolytes, and demonstrate its utility in regimes where linear-response liquid-state theories fail due to strong non-linear coupling between electrostatics and steric cavity interactions.
	
	\section{Model and Theoretical Framework}
	
	Consider an asymmetric $\alpha_+:\alpha_-$ electrolyte formed by the complete dissociation of a neutral salt,
	\begin{equation}
		M_{\nu_+}^{\alpha_+} X_{\nu_-}^{\alpha_-} \rightleftharpoons \nu_+ M^{\alpha_+} + \nu_- X^{\alpha_-} \ ,
	\end{equation}
	where $\alpha_+$ and $\alpha_-$ denote the absolute ionic valencies. Global charge neutrality of the salt formula unit dictates that
	\begin{equation}
		\nu_+ \alpha_+ - \nu_- \alpha_- = 0 \ .
	\end{equation}
	In a macroscopic, charge-neutral bulk solution at salt concentration $c_b$, the stoichiometric individual ionic concentrations are given by $c_+ = \nu_+ c_b$ and $c_- = \nu_- c_b$.
	
	When a simulation domain is in open thermodynamic equilibrium with an external electrolyte reservoir, phase equilibrium requires the equality of single-ion electrochemical potentials,
	\begin{equation}
		\tilde{\mu}_\pm^{\text{cell}} = \tilde{\mu}_\pm^{\text{reservoir}} \ .
	\end{equation}
	To establish a rigorous single-ion Grand Canonical Monte Carlo (GCMC) framework, all electrostatic potentials are defined relative to the container walls of the reference reservoir, which we set as our $0$-potential reference state. By placing a surface charge density on these reservoir walls, the inner electrostatic (Galvani) potential of the bulk reservoir liquid is shifted to $+\psi_R$ with respect to the walls. Tuning this wall surface charge enables us to set the single-ion electrochemical potentials in the bulk reservoir to a chosen activity level $c_r$:
	\begin{equation}
		\beta \tilde{\mu}_\pm^{\text{reservoir}} = \ln\left(\frac{c_r}{c^\ominus}\right) \ ,
	\end{equation}
	where $c^\ominus = 1\text{ M}$ is the standard reference concentration. Note that in this specific case we have adjusted $\psi_R$ so that the electrochemical potential of cations and anions become identical even for an asymmetric electrolyte.  
	
	Inside the periodic Ewald simulation cell, an entering ion experiences local interionic correlation and steric work $\mu_\pm^{\text{ex, local}}$, together with the electrostatic potential step across the macroscopic crystal boundary---the Bethe potential drop $q_\pm \langle \phi_B \rangle$---when crossing from the external reservoir into the simulation cell (see Appendix~\ref{sec:app_bethe}). Measured relative to the $0$-potential reservoir walls, the mean electrostatic potential inside the simulation box is $\langle \phi_B + \psi_R \rangle$. The total single-ion electrochemical potential inside the cell is therefore written as
	\begin{equation}
		\beta \tilde{\mu}_\pm^{\text{cell}} = \ln\left(\frac{\langle c_\pm \rangle}{c^\ominus}\right) + \beta \mu_\pm^{\text{ex, local}} + \beta q_\pm \langle \phi_B + \psi_R \rangle \ .
	\end{equation}
	
	Equating the single-ion electrochemical potentials between the reservoir and the simulation cell ($\tilde{\mu}_\pm^{\text{reservoir}} = \tilde{\mu}_\pm^{\text{cell}}$) yields the fundamental phase equilibrium condition:
	\begin{equation}\label{eq:phase_eq}
		\ln\left(\frac{c_r}{c^\ominus}\right) = \ln\left(\frac{\langle c_\pm \rangle}{c^\ominus}\right) + \beta \mu_\pm^{\text{ex, local}} + \beta q_\pm \langle \phi_B + \psi_R \rangle \ ,
	\end{equation}
	which directly isolates the local ionic excess chemical potential in terms of the combined potential average $\langle \phi_B + \psi_R \rangle$:
	\begin{equation}\label{eq:mulocal_isolated}
		\beta \mu_\pm^{\text{ex, local}} = \ln\left(\frac{c_r}{\langle c_\pm \rangle}\right) - \beta q_\pm \langle \phi_B + \psi_R \rangle \ .
	\end{equation}
	
	The complete physical excess chemical potential $\mu_\pm^{\text{ex}}$ of an ion inside the simulation box combines the local correlation contribution and the macroscopic boundary potential term:
	\begin{equation}
		\beta \mu_\pm^{\text{ex}} = \beta \mu_\pm^{\text{ex, local}} + \beta q_\pm \langle \phi_B \rangle \ .
	\end{equation}
	As shown in Appendix~\ref{sec:app_bethe}, under standard periodic boundary conditions (PBC) for spatially homogeneous bulk electrolytes, $\langle \phi_B \rangle = 0$ identically, so that $\mu_\pm^{\text{ex}} = \mu_\pm^{\text{ex, local}}$ and
	\begin{equation}\label{eq:muex_isolated}
		\beta \mu_\pm^{\text{ex}} = \ln\left(\frac{c_r}{\langle c_\pm \rangle}\right) - \beta q_\pm \langle \phi_B + \psi_R \rangle \ .
	\end{equation}
	Evaluating $\mu_\pm^{\text{ex, local}}$ via the combined average $\langle \phi_B + \psi_R \rangle$ before applying $\langle \phi_B \rangle = 0$ is statistically superior to evaluating $\langle \psi_R \rangle$ alone: because the GCMC acceptance probabilities depend strictly on the sum $\phi_B + \psi_R$, the dynamic feedback on $\psi_R$ directly compensates instantaneous fluctuations in $\phi_B$, rendering $\phi_B$ and $\psi_R$ strongly anticorrelated and substantially reducing the statistical variance of $\langle \phi_B + \psi_R \rangle$.
	
	In an open simulation domain, an arbitrary choice for the potential offset $\psi_R$ creates an energetic imbalance, causing the cell to absorb an excess of one ionic species over another and resulting in a net uncompensated charge $Q_t = \sum_i q_i \ne 0$. While 3D periodic Ewald summation with tinfoil boundary conditions automatically neutralizes $Q_t$ via an implicit, uniform background charge, a simulation box with $\langle Q_t \rangle \ne 0$ remains an unphysical non-neutral artifact.
	
	To recover the true physical bulk state, the system must satisfy macroscopic charge neutrality on average, $\langle Q_t \rangle = 0$. This requires selecting the unique potential alignment $\psi_R$ under which the average ionic populations inside the cell obey strict stoichiometry, $\alpha_+ \langle c_+ \rangle - \alpha_- \langle c_- \rangle = 0$. Under this potential condition, the simulation cell represents a true macroscopic bulk electrolyte at equilibrium salt concentration $c_b = \langle c_+ \rangle / \nu_+ = \langle c_- \rangle / \nu_-$.
	
	\section{Simulation Protocol and Implementation}
	
	We consider a cubic simulation cell of volume $V = L^3$ containing $N_+$ cations of charge $q_+ = \alpha_+ e$ and radius $r_+$, and $N_-$ anions of charge $q_- = -\alpha_- e$ and radius $r_-$, immersed in a continuous medium of permittivity $\varepsilon$. The ions interact via hard-core repulsion at contact distances $r_i + r_j$ and via long-range Coulomb forces.
	
	Replicating the simulation box periodically in three dimensions, the total electrostatic energy $U$ evaluated via Ewald summation is given by~\cite{levin2023new}
	\begin{eqnarray}\label{ener}
		U &=& \sum_{{\bf n}'=-\infty}^{\infty}\frac{2\pi}{\varepsilon L^3 |\pmb k|^2}\exp\left(-\frac{|\pmb k|^2}{4\kappa_e^2}\right)\left[A(\pmb k)^2 + B(\pmb k)^2\right] \nonumber \\
		& & + \frac{1}{2}\sum_{i \ne j}^N \frac{q_i q_j \erfc(\kappa_e|\pmb r_i - \pmb r_j|)}{\varepsilon|\pmb r_i - \pmb r_j|} \nonumber \\
		& & - \sum_{i=1}^N \frac{q_i^2 \kappa_e}{\varepsilon \sqrt{\pi}} - \frac{\pi Q_t^2}{2\varepsilon L^3 \kappa_e^2}+ \frac{2 \pi \pmb M^2}{3 \varepsilon L^3 }\ ,
	\end{eqnarray}
	where $N = N_+ + N_-$ is the total particle number, $Q_t = \sum_{i=1}^N q_i$ is the net charge in the box, and the structural factors are
	\begin{eqnarray}
		&&A(\pmb k) = \sum_{i=1}^N q_i \cos(\pmb k \cdot \pmb r_i)  \nonumber\\
		&&B(\pmb k) = -\sum_{i=1}^N q_i \sin(\pmb k \cdot \pmb r_i) \nonumber\\
		&&{\pmb M}=\sum_{i=1}^N q_i \pmb r_i \ .
	\end{eqnarray}
	The reciprocal lattice vectors are $\pmb k = (2\pi/L)\pmb n$ with integer components $\pmb n = (n_x, n_y, n_z)$, and the prime indicates the exclusion of the zero vector $\pmb n = (0,0,0)$. The damping parameter is set to $\kappa_e = 5/L$. The $Q_t^2$ term in Eq.~\eqref{ener} accounts for the self-energy of the uniform compensating background charge that neutralizes non-zero net charge fluctuations ($Q_t \ne 0$) during single-ion insertion and deletion moves. The $\pmb M^2$ contribution arises from the $k=0$ term in the reciprocal Ewald summation. Physically, it represents the contribution to the energy arising from the ``bound'' charge at the surface of a macroscopic Ewald crystal, with surface charge density $\sigma_b \sim {\pmb M} \cdot {\pmb n}$, where $\pmb n$ is the outward normal to the surface of a macroscopic spherical Ewald crystal~\cite{levin2023new}. If the simulation cell has a non-zero trace of the quadrupole moment of the charge distribution, in addition to the bound charges at the surface of the Ewald crystal it will also have a dipole layer.
	Following Ref.~\cite{levin2023new}, the modified Bethe potential $\phi_B$ inside the simulation cell carrying an instantaneous net uncompensated charge $Q_t$ and a uniform neutralizing background is given by:
	\begin{equation}\label{eq:phiB_def}
		\phi_B = -\frac{2\pi}{3\varepsilon V}\sum_{i=1}^N q_i r_i^2 + \frac{\pi Q_t}{6\varepsilon L} \ ,
	\end{equation}
	where $\mathbf{r}_i \in [-L/2, L/2)^3$ are the ionic coordinates inside the periodic simulation box.
	We note that both the Bethe potential and the $\pmb M^2$ term in the energy come from the singular $k=0$ term in the reciprocal Ewald summation and vanish for tinfoil boundary conditions. It is also important to note that the Bethe potential does not directly contribute to the electrostatic energy of ions inside the cell; nevertheless, when an ion enters or exits the cell it has to do work against this potential. The simulation has two types of moves: grand-canonical insertion/deletion and canonical displacement of ions. During canonical moves, we imagine that the Ewald crystal is wrapped in tinfoil; the induced image charges then completely cancel the $\pmb M^2$ term in the energy as well as the Bethe potential, so canonical displacements are governed purely by the tinfoil Ewald energy. On the other hand, during insertion and deletion moves, the Ewald crystal must be unwrapped from the tinfoil so that an ion entering or leaving the simulation cell traverses the boundary layer of the crystal and feels the potential produced by both the bound surface charge $\sigma_b$ and the dipole layer. Consequently, when evaluating $\Delta U_{\text{Ewald}} = U_{\text{new}} - U_{\text{old}}$ for GCMC insertion and deletion moves, all pairwise, self-energy, net-charge background, and macroscopic surface-dipole ($\pmb M^2$) terms in Eq.~\eqref{ener} are included, while the dipole-layer potential $\phi_B$ enters explicitly in the acceptance criteria.
	
	The structural factors $A(\pmb k)$ and $B(\pmb k)$ depend on the coordinates of individual ions. Consequently, when an ion $i$ is inserted, deleted, or displaced, updating $A(\pmb k)$ and $B(\pmb k)$ requires modifying only the specific trigonometric terms corresponding to that single particle, rather than recalculating the full summation over all $N$ ions. This incremental $O(1)$ update renders the reciprocal-space energy calculation exceptionally fast during single-ion GCMC trial moves, further enhancing the overall computational efficiency of the algorithm.
	
	Single-ion trial moves are proposed according to the salt stoichiometry. Insertion attempts select a cation with probability $p_a(+) = \nu_+ / \nu_t$ or an anion with probability $p_a(-) = \nu_- / \nu_t$, where $\nu_t = \nu_+ + \nu_-$, generating the trial position $\mathbf{r}$ uniformly at random within the volume $V$. For deletion moves, an existing particle is chosen uniformly at random among all $N = N_+ + N_-$ particles currently present in the simulation cell, giving a selection probability $1/N$ for any specific ion.
	
	To drive the simulation cell toward macroscopic charge neutrality while enforcing thermodynamic equilibrium with the reservoir at reference activity $c_r$, the single-ion acceptance probabilities incorporating both the modified Bethe boundary potential $\phi_B$ and the dynamic potential shift $\psi_R$ are given by:
	\begin{equation}
		P_a(\pm) = \min\left[1, \frac{V c_r}{N_\pm + 1} e^{-\beta \Delta U_{\text{Ewald}} - \beta q_\pm \left[\phi_B + \psi_R\right]}\right] \ ,
	\end{equation}
	for inserting an ion of species $\pm$, and
	\begin{equation}
		P_d(\pm) = \min\left[1, \frac{N_\pm}{V c_r} e^{-\beta \Delta U_{\text{Ewald}} + \beta q_\pm \left[\phi_B + \psi_R\right]}\right] \ ,
	\end{equation}
	for removing an ion of species $\pm$.
	
	Here, $\phi_B$ is the Bethe boundary potential determined strictly by the $N$ background ions inside the cell, excluding the trial ion entering or leaving the system. For an insertion move ($N \to N+1$), $\phi_B$ is evaluated using the $N$ ions currently present in the cell prior to insertion. For a deletion move ($N+1 \to N$), $\phi_B$ is evaluated using the $N$ ions left behind after the deleted ion is removed. In both cases, the entering or leaving ion does not contribute to $\phi_B$. This ensures that the boundary work for insertion, $+q_\pm \phi_B$, is identically equal and opposite to the boundary work for deletion, $-q_\pm \phi_B$, rigorously preserving detailed balance. Note also that even though the ensemble averages of both the Bethe potential and the cell dipole moment vanish under PBC for a spatially homogeneous and isotropic bulk electrolyte ($\langle \phi_B \rangle = 0$ and $\langle \mathbf{M} \rangle = \mathbf{0}$, see Appendix~\ref{sec:app_bethe}), instantaneously within a given configuration there is both a non-zero Bethe potential $\phi_B$ and a finite dipole moment $\mathbf{M}$ inside the cell, so the ion crossing the unwrapped boundary feels these instantaneous contributions.

	During the simulation, the uniform potential shift $\psi_R$ is updated iteratively following each trial move to penalize any instant net charge accumulation $Q_t$:
	\begin{equation}
		\psi_R^{(n+1)} = \psi_R^{(n)} + \xi Q_t^{(n)}/N \ ,
		\label{eq:psi}
	\end{equation}
	where $\xi = 0.001 \ k_B T / e$ is a relaxation parameter, and the initial value is set to $\psi_R^{(0)} = 0$. In production, $10^6$ trial movements are typically performed for initial equilibration, followed by data collection over $10^4$ configurations sampled every $10^3$ ionic moves to obtain ensemble averages of the concentrations $\langle c_\pm \rangle$ and the combined electrostatic potential $\langle \phi_B + \psi_R \rangle$.
	
	Once the simulation reaches steady-state equilibrium, the dynamic update rule guarantees that the average net charge vanishes, $\langle Q_t \rangle = 0$. In this state, the ionic populations inside the box strictly obey the stoichiometric ratio dictated by charge neutrality, $\alpha_+ \langle c_+ \rangle - \alpha_- \langle c_- \rangle = 0$. Consequently, the simulation cell represents a true macroscopic bulk electrolyte solution at an equilibrium salt concentration $c_b = \langle c_+ \rangle / \nu_+ = \langle c_- \rangle / \nu_-$.
	
	The Monte Carlo update rules combine stoichiometric insertion proposals ($p_a(\pm) = \nu_\pm / \nu_t$) with uniform global deletion proposals ($1/N$). At instantaneous charge neutrality ($Q_t = 0$), the ionic populations strictly obey $N_+ / N_- = \nu_+ / \nu_-$, which algebraically guarantees that the probability of selecting an ion of species $\pm$ for deletion, $N_\pm / N$, identically equals its insertion proposal probability, $\nu_\pm / \nu_t$. Any transient fluctuation away from neutrality ($Q_t \ne 0$) introduces a momentary proposal asymmetry; however, these charge fluctuations are dynamically regulated by the iterative update of the uniform potential shift, Eq.~\eqref{eq:psi}. During the transient equilibration phase, this non-Markovian feedback deliberately breaks detailed balance, acting as a deterministic gradient descent superimposed on the stochastic Monte Carlo chain to forcefully drive the system toward macroscopic electroneutrality ($\langle Q_t \rangle = 0$). Crucially, once this equilibrium state is achieved, detailed balance is effectively restored. With the system fluctuating symmetrically around $Q_t = 0$, the ensemble average $\langle N_\pm / N \rangle$ rigorously converges to $\nu_\pm / \nu_t$, neutralizing the structural proposal asymmetry, while $\langle \phi_B + \psi_R \rangle$ anchors at a stable equilibrium value. In this production phase, the residual microscopic energetic mismatch between forward and reverse acceptance probabilities scales as $\exp\left[-\beta q_\pm \xi Q_t / N\right]$. By maintaining a sufficiently small relaxation parameter ($\xi = 0.001 \ k_B T / e$), this transient discrepancy remains negligible compared to thermal fluctuations and rigorously time-averages to zero over the sampled ensemble.
	
	Thermodynamically, phase equilibrium between the reference reservoir and the neutral simulation box requires the equality of the single-ion electrochemical potentials. Equating $\tilde{\mu}_\pm^{\text{reservoir}}$ and $\tilde{\mu}_\pm^{\text{cell}}$ isolates the local excess chemical potential $\mu_\pm^{\text{ex, local}}$ in terms of the sampled bulk concentrations $\langle c_\pm \rangle$ and the converged average $\langle \phi_B + \psi_R \rangle$; invoking $\langle \phi_B \rangle = 0$ then yields the complete individual ionic excess chemical potentials $\mu_\pm^{\text{ex}}$:
	\begin{equation}\label{eq:muex_final}
		\beta \mu_\pm^{\text{ex}} = \beta \mu_\pm^{\text{ex, local}} = \ln\left(\frac{c_r}{\langle c_\pm \rangle}\right) - \beta q_\pm \langle \phi_B + \psi_R \rangle \ .
	\end{equation}
	In order to maintain $\langle Q_t \rangle = 0$, $\phi_B$ and $\psi_R$ fluctuate in opposition (are anti-correlated), so sampling the combined average $\langle \phi_B + \psi_R \rangle$ achieves significantly faster convergence than averaging $\psi_R$ alone. Furthermore, because both cations and anions are sampled simultaneously during the single-ion GCMC moves, a single simulation run at a fixed reference activity $c_r$ rapidly yields the complete pair of individual excess chemical potentials $\{\mu_+^{\text{ex}}, \mu_-^{\text{ex}}\}$ corresponding to the equilibrium bulk concentration $c_b$.
	
	\section{Results and Discussion}
	
\begin{figure}[t]
		\begin{center}
			\includegraphics[width=0.48\textwidth]{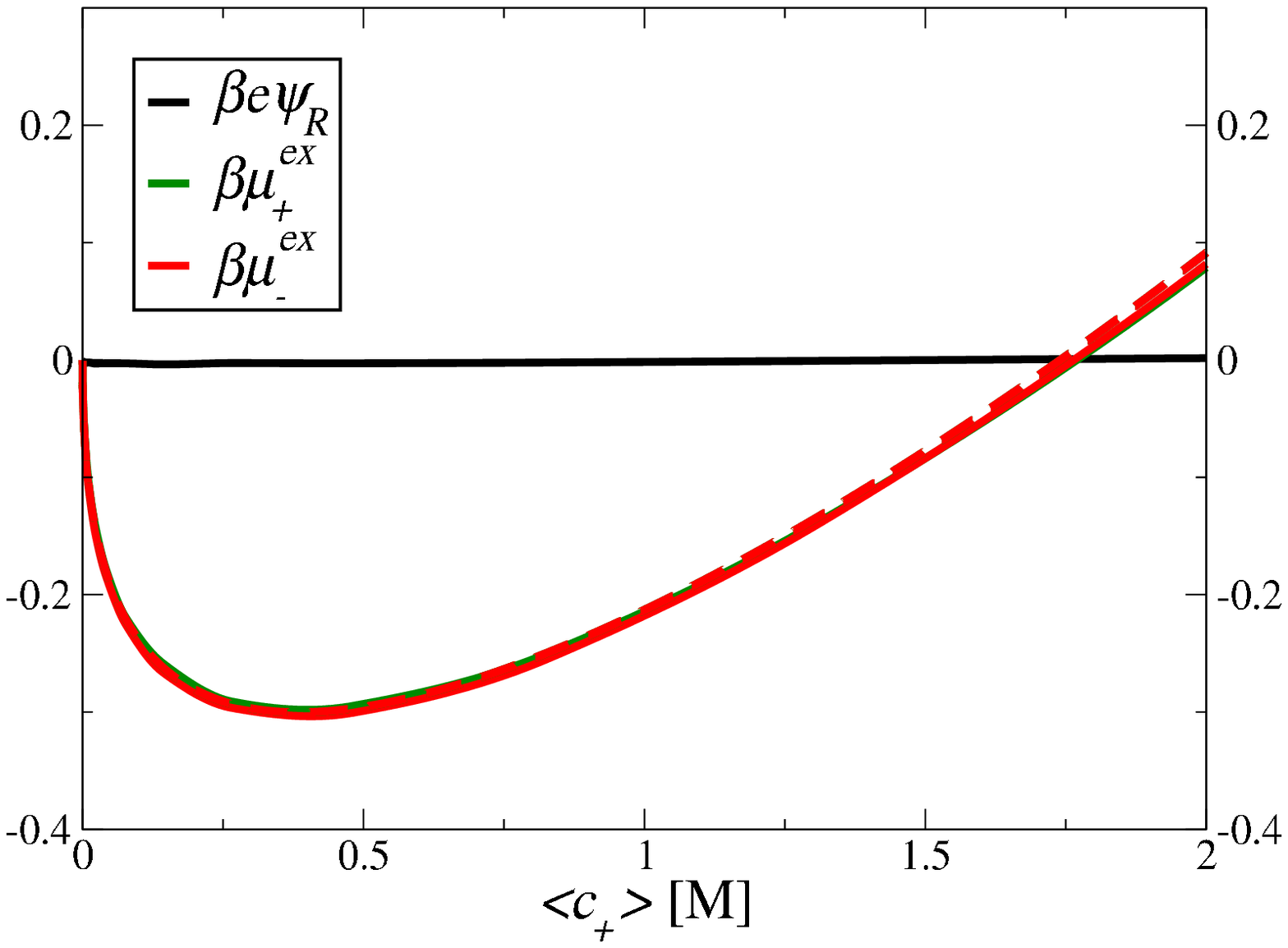}
			\caption{Individual excess chemical potentials $\beta \mu_+^{\text{ex}}$ (monovalent cations) and $\beta \mu_-^{\text{ex}}$ (monovalent anions) for a $1:1$ electrolyte with size-symmetric ions ($r_+ = 2\text{ \AA}$, $r_- = 2\text{ \AA}$). Solid lines represent Donnan GCMC simulation results and dashed lines denote MSA theory. The solid black line shows the average reservoir potential $\beta e \psi_R$. For symmetric electrolyte the chemical potentials of anions and cations are identical, the curves overlap.  Furthermore, the MSA is so accurate that it is indistinguishable from simulation results.}
			\label{fig:chem1122}
		\end{center}
	\end{figure}

\begin{figure}[t]
		\begin{center}
			\includegraphics[width=0.48\textwidth]{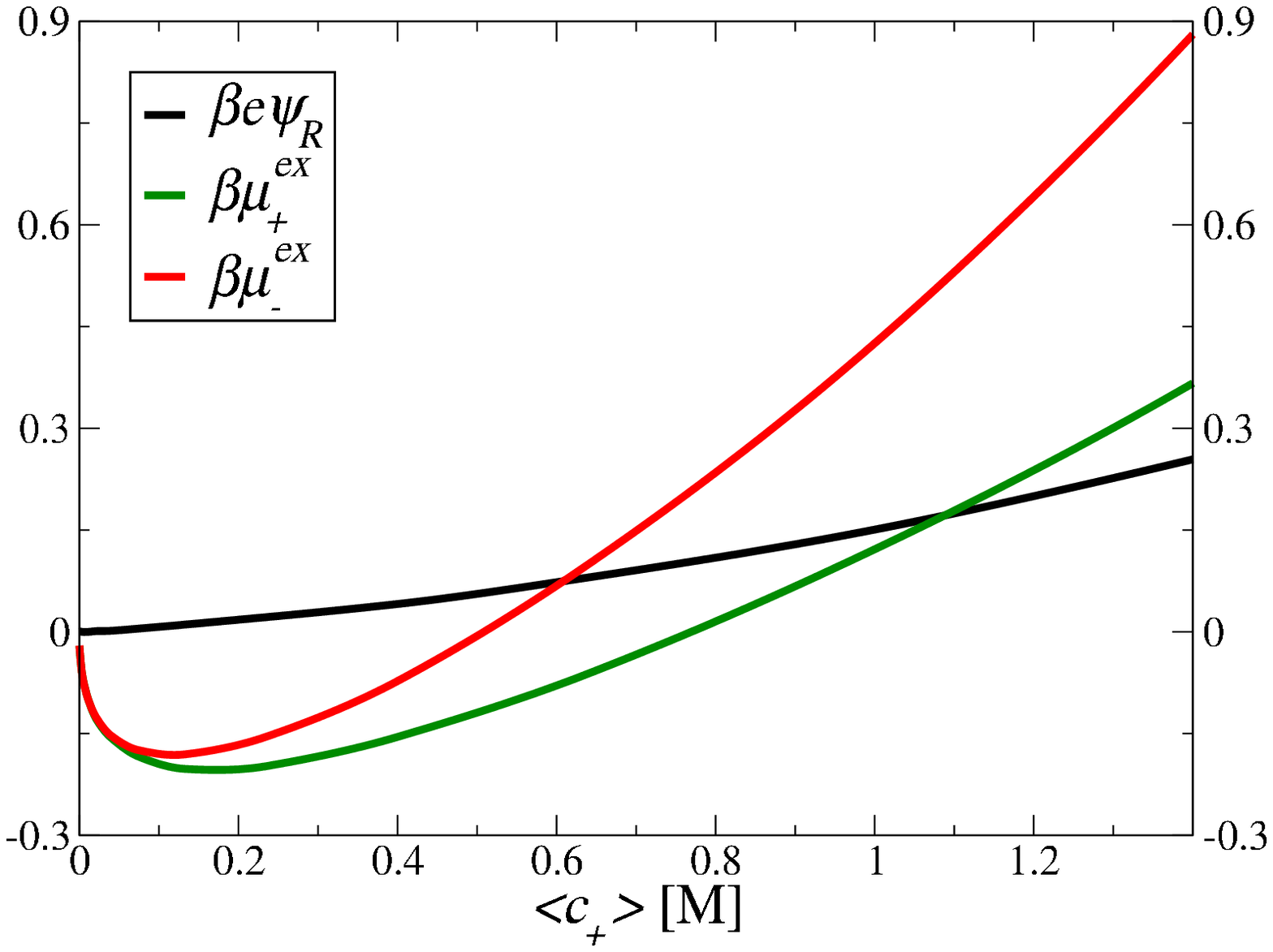}
			\caption{Individual excess chemical potentials and average reservoir potential $\beta e \psi_R$ obtained from Donnan GCMC simulations for a $1:1$ electrolyte with size-asymmetric ions ($r_+ = 2\text{ \AA}$, $r_- = 3\text{ \AA}$).}
			\label{fig:chem1124}
		\end{center}
	\end{figure}

\begin{figure}[t]
		\begin{center}
			\includegraphics[width=0.45\textwidth]{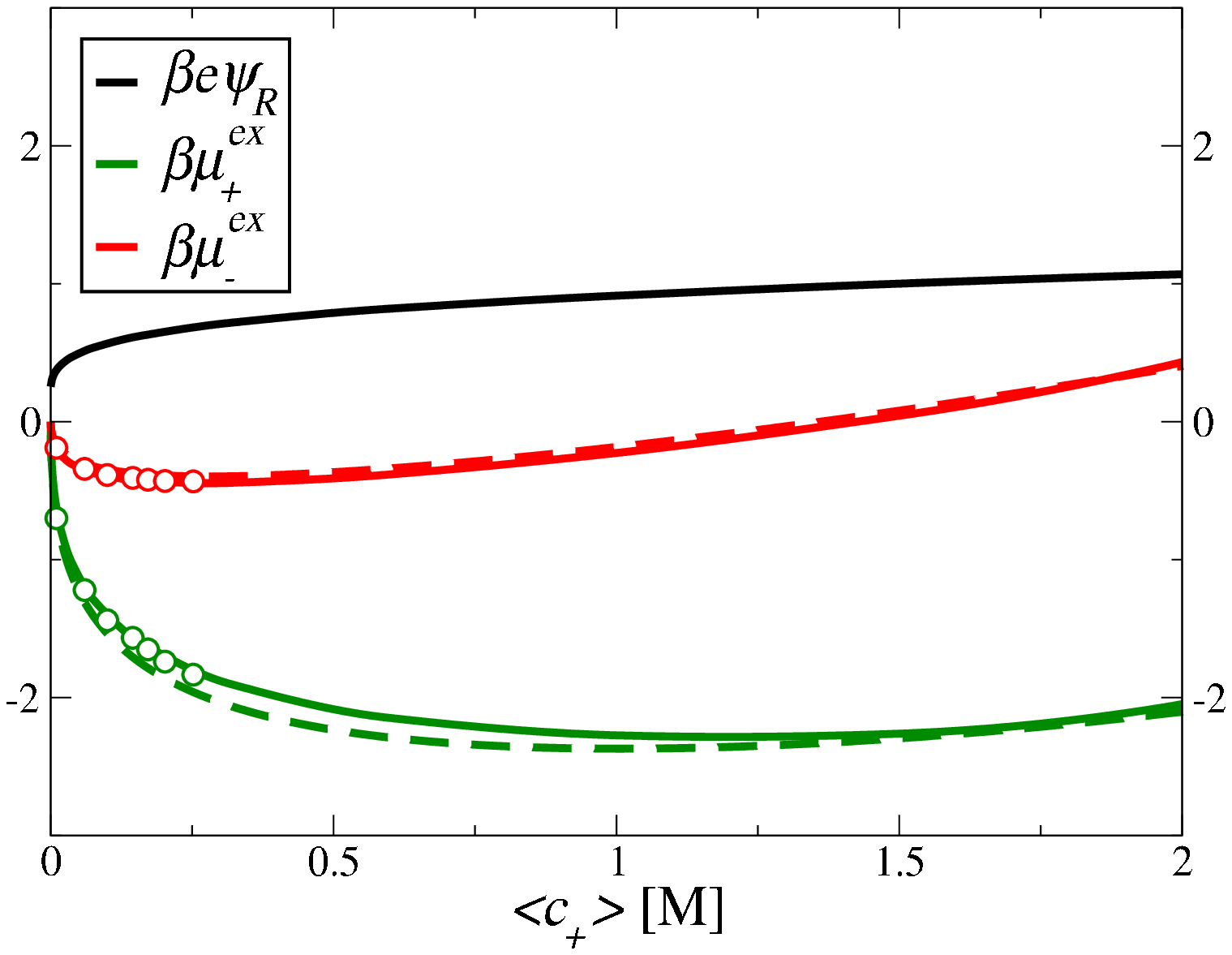}
			\caption{Individual excess chemical potentials $\beta \mu_+^{\text{ex}}$ (multivalent cations) and $\beta \mu_-^{\text{ex}}$ (monovalent anions) for a $2:1$ electrolyte with size-symmetric ions ($r_+ = 2\text{ \AA}$, $r_- = 2\text{ \AA}$). Solid lines represent Donnan GCMC simulation results, dashed lines denote MSA theory, open markers show direct single-particle Widom insertion calculations, and the solid black line shows $\beta e \psi_R$.}
			\label{fig:chem2122}
		\end{center}
	\end{figure}

\begin{figure}[t]
		\begin{center}
			\includegraphics[width=0.45\textwidth]{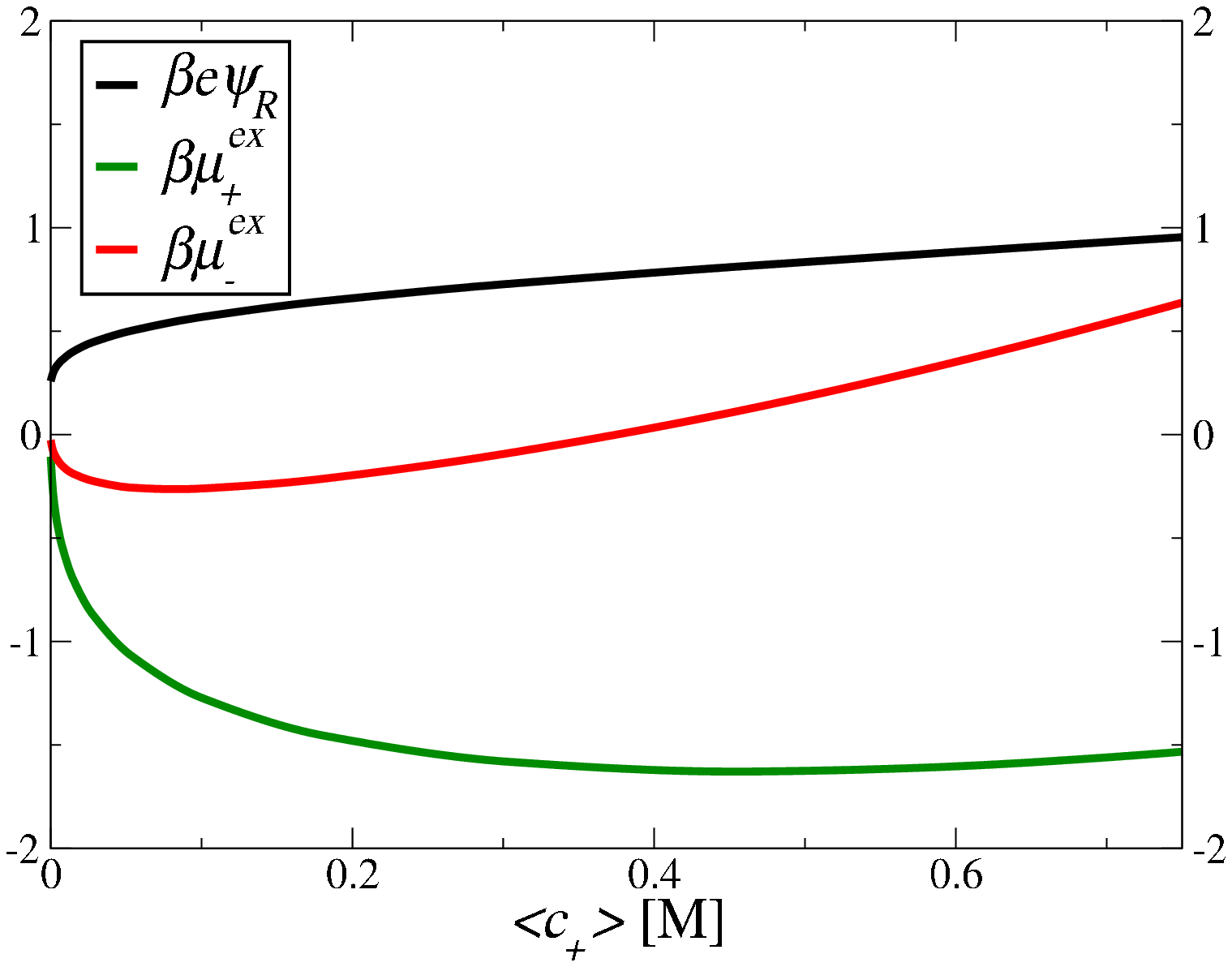}
			\caption{Individual excess chemical potentials and average reservoir potential $\beta e \psi_R$ obtained from Donnan GCMC simulations for a $2:1$ electrolyte with size-asymmetric ions ($r_+ = 2\text{ \AA}$, $r_- = 3\text{ \AA}$).}
			\label{fig:chem2124}
		\end{center}
	\end{figure}

\begin{figure}[t]
		\begin{center}
			\includegraphics[width=0.45\textwidth]{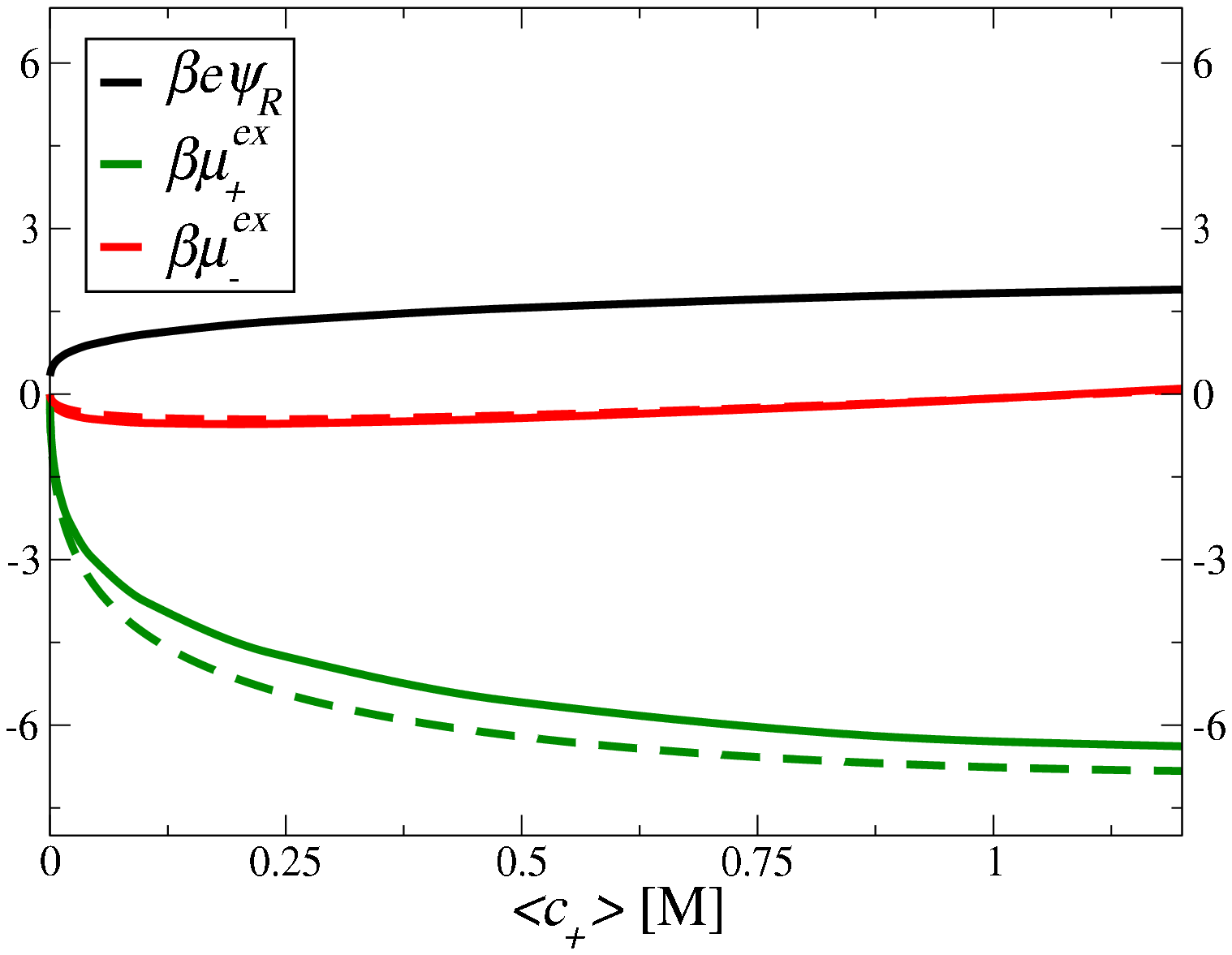}
			\caption{Individual excess chemical potentials for a $3:1$ size-symmetric electrolyte ($r_+ = 2\text{ \AA}$, $r_- = 2\text{ \AA}$). Dashed lines represent MSA theory, solid colored lines denote Donnan GCMC simulation data, and the solid black line shows $\beta e \psi_R$.}
			\label{fig:chem3122}
		\end{center}
	\end{figure}

\begin{figure}[t]
		\begin{center}
			\includegraphics[width=0.45\textwidth]{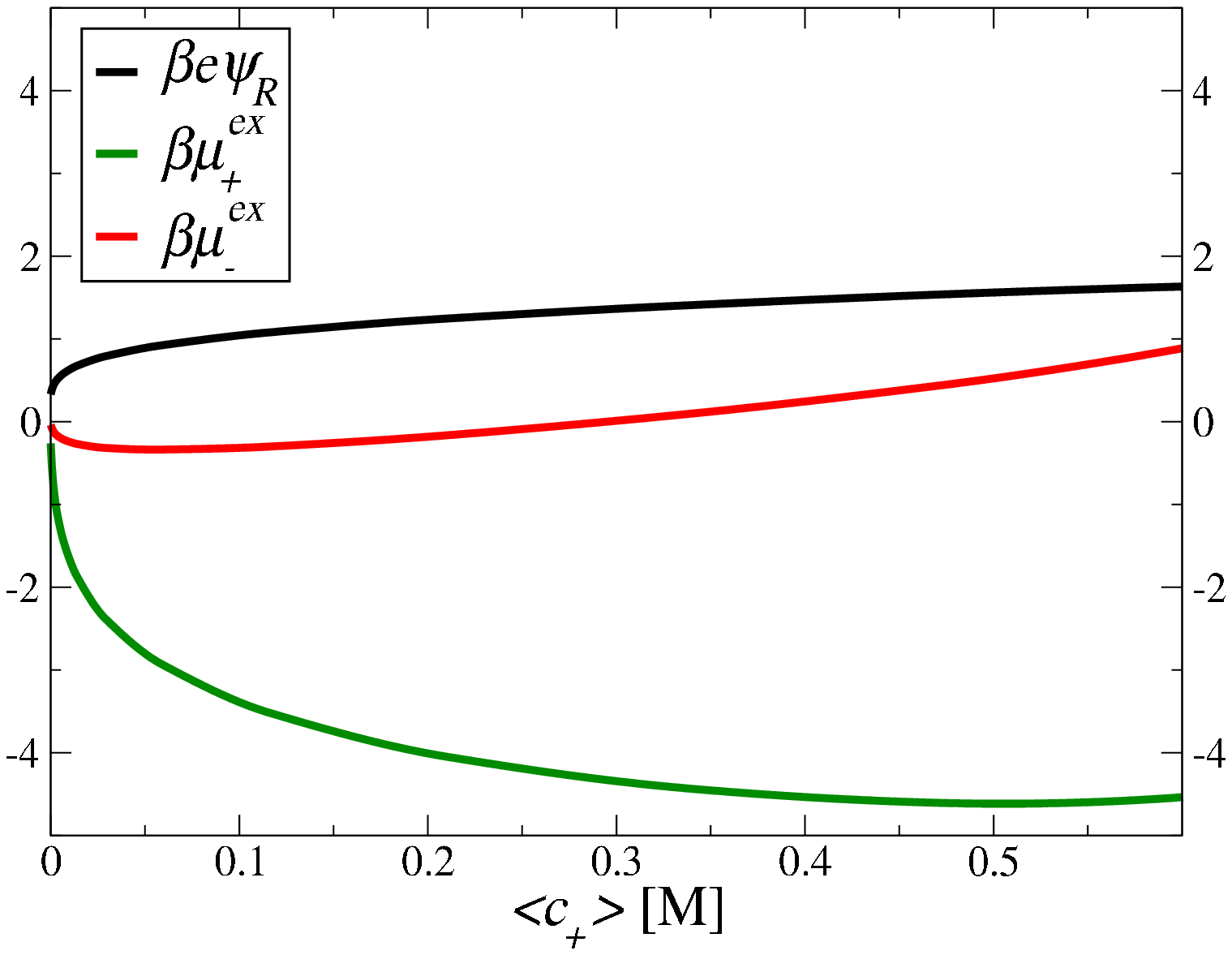}
			\caption{Individual excess chemical potentials and average reservoir potential $\beta e \psi_R$ obtained from Donnan GCMC simulations for a $3:1$ size-asymmetric electrolyte ($r_+ = 2\text{ \AA}$, $r_- = 3\text{ \AA}$).}
			\label{fig:chem3124}
		\end{center}
	\end{figure}

	To demonstrate the accuracy and numerical efficiency of the Donnan GCMC method, we investigate aqueous electrolyte solutions of varying valency stoichiometries ($1:1$, $2:1$, and $3:1$) and ionic size asymmetries. In all simulations, the solvent is modeled as a continuous dielectric medium with relative permittivity $\varepsilon = 78.4$ at room temperature $T = 298.15\text{ K}$, corresponding to a Bjerrum length $l_B = \beta e^2 / \varepsilon = 7.14\text{ \AA}$. Because $\langle \phi_B \rangle = 0$ under PBC, the converged equilibrium average $\langle \phi_B + \psi_R \rangle$ entering Eq.~\eqref{eq:muex_final} equals $\langle \psi_R \rangle$, which is plotted alongside the excess chemical potentials in Figs.~\ref{fig:chem1122}--\ref{fig:chem3124} (denoted simply as $\beta e \psi_R$).
	
	The theoretical benchmark for comparison is the Mean Spherical Approximation (MSA)~\cite{ho1988mean}, which is very accurate for $1:1$ symmetric electrolytes in the weak coupling limit relevant for water at room temperature. Within the MSA framework, the individual ionic excess chemical potential $\mu_i^{\text{MSA}}$ is expressed as the sum of an electrostatic correlation term and a hard-core excluded-volume (cavity) contribution:
	\begin{equation}
		\beta \mu_i^{\text{MSA}} = \beta \mu_i^{\text{el}} + \beta \mu_i^{\text{hc}} \ .
	\end{equation}
	The electrostatic contribution for an ion of valence $\alpha_i$ and radius $r_i$ is given by
	\begin{equation}
		\beta \mu_i^{\text{el}} = -\frac{\alpha_i^2 l_B \Gamma}{1 + 2\Gamma r_i} \ ,
	\end{equation}
	where for size-symmetric ions ($r_+ = r_- = r$), the screening parameter simplifies to $\Gamma = (-1 + \sqrt{1 + 4\kappa r}) / (4r)$, with $\kappa = \sqrt{4\pi l_B \sum_j \alpha_j^2 c_j}$ being the inverse Debye length.
	
	For the hard-core cavity contribution in size-symmetric systems ($r_+ = r_- = r$), we employ the Carnahan-Starling excess chemical potential for hard spheres:
	\begin{equation}
		\beta \mu^{\text{hc}}_i = \frac{8\phi - 9\phi^2 + 3\phi^3}{(1 - \phi)^3} \ ,
	\end{equation}
	where $\phi = \frac{4\pi}{3} r^3 c_t$ is the total ionic packing fraction, with $c_t = c_+ + c_-$.
	
	Figure~\ref{fig:chem1122} shows the individual excess chemical potentials for a $1:1$ electrolyte with size-symmetric ions ($r_+ = r_- = 2\text{ \AA}$). Symmetry guarantees $\mu_+^{\text{ex}} = \mu_-^{\text{ex}}$, and the Donnan GCMC results agree perfectly with the benchmark MSA theory across the entire concentration range.

	In Fig.~\ref{fig:chem1124}, we present the chemical potentials for a $1:1$ size-asymmetric ($r_+ = 2\text{ \AA}$, $r_- = 3\text{ \AA}$) electrolyte. When the anion radius is increased to $r_- = 3\text{ \AA}$ in Fig.~\ref{fig:chem1124}, stronger hard-core excluded-volume repulsion elevates $\mu_-^{\text{ex}}$ significantly above $\mu_+^{\text{ex}}$ at moderate-to-high salt concentrations. Unfortunately, MSA is known to fail for size-asymmetric electrolytes. For this reason, we do not plot it in Fig.~\ref{fig:chem1124}.
	
	Fig.~\ref{fig:chem2122} shows the individual excess chemical potentials for a $2:1$ electrolyte with size-symmetric ions ($r_+ = 2\text{ \AA}$ and $r_- = 2\text{ \AA}$). The Donnan GCMC results show excellent agreement with direct single-particle Widom insertion calculations from Ref.~\cite{bakhshandeh2022widom}. Surprisingly, we find that MSA also agrees reasonably well with simulation across the entire concentration range up to $2\text{ M}$, even though $2:1$ electrolytes cross into an intermediate coupling regime. Notably, while the excess chemical potential of the divalent cations $\mu_+^{\text{ex}}$ exhibits strong negative electrostatic corrections due to multivalent correlations, the monovalent anions $\mu_-^{\text{ex}}$ display a milder decrease. The seamless match with Widom insertion simulations confirms the thermodynamic rigor of our dynamic potential protocol, while achieving numerical convergence in a fraction of the computational time. Fig.~\ref{fig:chem2124} presents the simulation results for a $2:1$ salt with $r_+ = 2\text{ \AA}$ and larger anions $r_- = 3\text{ \AA}$. It is shown that a slight change in ionic size reflects in a huge difference in the excess chemical potential profiles compared to the symmetrical case.  
	
	For the $3:1$ symmetric electrolyte ($r_+ = 2\text{ \AA}$, $r_- = 2\text{ \AA}$) shown in Fig.~\ref{fig:chem3122}, strong electrostatic correlations around the trivalent cations dominate the low-concentration regime, causing a steep drop in $\mu_+^{\text{ex}}$. Here, the agreement between Donnan GCMC and MSA theory remains good up to concentrations around $50\text{ mM}$, beyond which deviations steadily grow as multivalent ion correlations strengthen. Finally, Fig.~\ref{fig:chem3124} displays the results for the $3:1$ size-asymmetric electrolyte ($r_+ = 2\text{ \AA}$, $r_- = 3\text{ \AA}$), illustrating the smooth performance and rapid statistical convergence of the Donnan GCMC protocol even in highly asymmetric regimes.
	
	\section{Conclusions}
	
	In this work, we have developed a Donnan Grand Canonical Monte Carlo simulation methodology using fast 3D Ewald summation to calculate individual ionic excess chemical potentials in electrolyte solutions. By introducing a dynamic inner electrostatic (Galvani) potential shift $\psi_R$ that dynamically penalizes uncompensated net charge fluctuations, the simulation cell is driven to satisfy macroscopic charge neutrality on average ($\langle Q_t \rangle = 0$). This self-consistent potential alignment enables active, single-ion GCMC insertion and deletion moves without triggering divergent background potential terms or violating thermodynamic consistency.

	We validated the protocol against direct single-particle Widom insertion calculations (with Ewald summation) and Mean Spherical Approximation (MSA) theory for size-symmetric $1:1$ and $2:1$ electrolytes. Excellent agreement with MSA was observed up to molar concentrations for symmetric $1:1$ and $2:1$ salts, while deviations emerge above $50\text{ mM}$ for $3:1$ salts due to strong electrostatic correlations. The simulation method allows us to obtain the chemical potentials of highly size-asymmetric electrolytes ($r_+ \ne r_-$), for which due to non-linear coupling between electrostatic screening and hard-core steric cavity interactions, no accurate liquid-state theory is available.
	
	Unlike standard neutral-pair GCMC algorithms---which suffer from severe phase-space sampling bottlenecks in dense or confined environments---and single-particle Widom insertions---which require extensive sampling of the phase space and separate, computationally expensive virtual insertions for each species---the Donnan GCMC protocol simultaneously yields the complete set of individual excess chemical potentials $\{\mu_+^{\text{ex}}, \mu_-^{\text{ex}}\}$ and the corresponding physical bulk concentration $c_b$ in a single simulation run with rapid numerical convergence. Explicit comparison shows that convergence of the Donnan GCMC algorithm is an order of magnitude faster than that of the Widom insertion method. Furthermore, in a single simulation we obtain the chemical potential of both the anion and the cation, while Widom insertion requires two separate simulations. 
	
	Beyond bulk solution thermodynamics, the framework established here provides the exact single-ion excess chemical potentials required to set up thermodynamic equilibrium in grand canonical simulations of inhomogeneous and nano-confined systems---such as supercapacitor porous electrodes, biological channels, and clay minerals---bypassing the prohibitive computational overhead of simulating complete, multi-component electrochemical cells.
	
	\begin{acknowledgments}
		This work was partially supported by CNPq under the grant 303310/2025-1.
	\end{acknowledgments}
    
	\section{Data and Software Availability}
The data and the underlying theoretical framework are fully contained in the published article and its appendix.

\appendix
\section{Bethe Surface Potential and Detailed Balance in Periodic Ewald Simulations}\label{sec:app_bethe}

In this Appendix, we show why the ensemble average of the Bethe surface potential vanishes ($\langle \phi_B \rangle = 0$) for spatially homogeneous bulk electrolytes under periodic boundary conditions (PBC), and outline the on-demand computational implementation and detailed balance conditions during Monte Carlo trial moves.

\subsection{Vanishing Average Bethe Potential under Periodic Boundary Conditions}

Consider a cubic simulation cell of volume $V = L^3$ centered on the origin, containing $N$ ions with charges $q_i$ and position vectors $\mathbf{r}_i = (x_i, y_i, z_i)$. Under 3D Ewald summation, the instantaneous Bethe potential is determined by the second radial moment of the charge distribution in the cell [Eq.~\eqref{eq:phiB_def}]:
\begin{equation}\label{eq:app_phiB_def}
	\phi_B = -\frac{2\pi}{3\varepsilon V}\sum_{i=1}^N q_i |\mathbf{r}_i|^2 + \frac{\pi Q_t}{6\varepsilon L} \ ,
\end{equation}
where $Q_t = \sum_{i=1}^N q_i$ is the instantaneous net charge in the box.

Under standard 3D periodic boundary conditions, particle coordinates reside within the central simulation box $-L/2 \le x_i, y_i, z_i < L/2$. For a spatially homogeneous bulk electrolyte, translational invariance dictates that the single-particle density profile $\langle \rho_s(\mathbf{r}) \rangle$ for each species $s \in \{+, -\}$ is uniform across the volume:
\begin{equation}
	\langle \rho_s(\mathbf{r}) \rangle = \frac{N_s}{V} \ .
\end{equation}

Evaluating the ensemble average of the second radial moment sum gives:
\begin{eqnarray}\label{eq:app_wrapped_integral}
	\left\langle \sum_{i=1}^N q_i |\mathbf{r}_i|^2 \right\rangle &=& \sum_{s=\pm} q_s \int_V |\mathbf{r}|^2 \langle \rho_s(\mathbf{r}) \rangle \, d^3r \nonumber \\
	&=& \left( \frac{1}{V} \int_V |\mathbf{r}|^2 \, d^3r \right) \sum_{s=\pm} q_s N_s \ .
\end{eqnarray}
The geometric integral over the cubic cell volume $V = L^3$ evaluates to
\begin{equation}
	C_{\text{box}} = \frac{1}{L^3} \int_{-L/2}^{L/2} \int_{-L/2}^{L/2} \int_{-L/2}^{L/2} (x^2 + y^2 + z^2) \, dx \, dy \, dz = \frac{L^2}{4} \ ,
\end{equation}
which yields
\begin{equation}
	\left\langle \sum_{i=1}^N q_i |\mathbf{r}_i|^2 \right\rangle = C_{\text{box}} \langle Q_t \rangle \ .
\end{equation}

Taking the ensemble average of Eq.~\eqref{eq:app_phiB_def} therefore yields:
\begin{eqnarray}
	\langle \phi_B \rangle &=& -\frac{2\pi}{3\varepsilon V} C_{\text{box}} \langle Q_t \rangle + \frac{\pi \langle Q_t \rangle}{6\varepsilon L} \nonumber \\
	&=& \left( -\frac{2\pi}{3\varepsilon V} \frac{L^2}{4} + \frac{\pi}{6\varepsilon L} \right) \langle Q_t \rangle = 0 \ .
\end{eqnarray}
Thus, under PBC, $\langle \phi_B \rangle = 0$ identically for any homogeneous bulk electrolyte, regardless of ionic valency or electrostatic coupling strength.

\subsection{Computational Implementation and Detailed Balance in Monte Carlo Simulations}

A critical distinction must be drawn between canonical spatial displacement moves and Grand Canonical Monte Carlo (GCMC) particle exchange moves regarding the boundary terms. Thermodynamically, during canonical displacement moves the simulation cell is embedded in conducting tinfoil boundary conditions, which sets both the macroscopic surface dipole term $\frac{2\pi \mathbf{M}^2}{3\varepsilon L^3}$ (arising from the $k=0$ reciprocal term) and the Bethe potential $\phi_B$ strictly to zero. Because the total particle number $N$ remains fixed and no charge crosses the macroscopic boundary, internal fluid sampling is governed purely by the tinfoil-shielded Ewald energy. Consequently, canonical displacements are accepted or rejected strictly according to the tinfoil Ewald energy difference, $P_{\text{acc}} = \min\left[1, \exp(-\beta \Delta U_{\text{tinfoil}})\right]$.

Conversely, during GCMC single-ion insertion and deletion trial moves, an ion explicitly crosses the macroscopic boundary of the infinite Ewald crystal. Conceptually, to transfer an ion between the external reference reservoir and the cell, the system is briefly unwrapped from its tinfoil sheath to account for the electrostatic work required to cross the macroscopic crystal interface. Even though $\langle \phi_B \rangle = 0$ and $\langle \mathbf{M} \rangle = \mathbf{0}$ on average under PBC for a spatially homogeneous and isotropic bulk electrolyte, instantaneously within a given configuration there is both a non-zero Bethe potential $\phi_B$ and a finite cell dipole moment $\mathbf{M} = \sum_{i=1}^N q_i \mathbf{r}_i$ inside the cell. During the boundary transfer, the ion feels these instantaneous boundary contributions: the change in the macroscopic surface-charge dipole energy $\Delta\left(\frac{2\pi \mathbf{M}^2}{3\varepsilon L^3}\right)$ is included in $\Delta U_{\text{Ewald}}$ [Eq.~\eqref{ener}], while the surface dipole-layer potential $\phi_B$ enters explicitly in the boundary crossing work $q_\pm (\phi_B + \psi_R)$ in $P_a(\pm)$ and $P_d(\pm)$. Once the particle exchange is completed, the cell is rewrapped in tinfoil to resume canonical equilibration.

To preserve strict detailed balance (microscopic reversibility), the Bethe boundary work associated with transferring an ion between the reservoir and the simulation cell must be evaluated using the $N$ background ions of the state without the entering or leaving particle. That is, for an insertion move ($N \to N+1$), $\phi_B$ is calculated using the $N$ ions in the cell before insertion. For a deletion move ($N+1 \to N$), $\phi_B$ is calculated using the $N$ ions left behind after removing the deleted particle. In both directions, the ion crossing the boundary does not contribute to $\phi_B$, ensuring $\Delta E_{\text{deletion}} = -\Delta E_{\text{insertion}}$.

Evaluating $\phi_B$ on-demand at the instant a GCMC trial move is proposed proceeds according to the following execution flow:
\begin{enumerate}
	\item \textbf{Compute Background Second Moment $S^{(N)}$ [$O(N)$]:} Using the periodic box coordinates $\mathbf{r}_i \in [-L/2, L/2)^3$ of the $N$ background ions (the $N$ ions currently in the cell for insertion, or the $N$ ions remaining after deletion), evaluate the second radial moment $S^{(N)} = \sum_{i=1}^N q_i |\mathbf{r}_i|^2$ and net background charge $Q_t^{(N)} = \sum_{i=1}^N q_i$.
	
	\item \textbf{Compute Background Bethe Potential $\phi_B$:} Substitute $S^{(N)}$ and $Q_t^{(N)}$ into Eq.~\eqref{eq:app_phiB_def}:
	\begin{equation}
		\phi_B = -\frac{2\pi}{3\varepsilon V} S^{(N)} + \frac{\pi Q_t^{(N)}}{6\varepsilon L} \ .
	\end{equation}
	
	\item \textbf{Evaluate Trial Energy Change and Acceptance:} For trial insertion of species $\pm$ at $\mathbf{r}_{\text{trial}}$, the net energy change entering $P_a(\pm)$ is:
	\begin{equation}
		\Delta E_{\text{ins}} = \Delta U_{\text{Ewald}} + q_\pm \left( \phi_B + \psi_R \right) \ .
	\end{equation}
	For trial deletion of ion $k$ of species $\pm$, the net energy change entering $P_d(\pm)$ is:
	\begin{equation}
		\Delta E_{\text{del}} = \Delta U_{\text{Ewald}} - q_\pm \left( \phi_B + \psi_R \right) \ .
	\end{equation}
	Here $\Delta U_{\text{Ewald}}$ includes all Ewald pairwise, self-energy, net-charge background, and macroscopic surface-dipole ($\mathbf{M}^2$) terms between the $N+1$ and $N$ states.
\end{enumerate}


	\bibliography{ref.bib}
	
\end{document}